\documentclass[9pt,conference,letter]{IEEEtran}

\IEEEoverridecommandlockouts
\usepackage{cite}
\usepackage{amsmath,amssymb,amsfonts}
\usepackage{algorithm}
\usepackage{algpseudocode}
\usepackage{graphicx}
\usepackage{textcomp}
\usepackage{xcolor}
\def\BibTeX{{\rm B\kern-.05em{\sc i\kern-.025em b}\kern-.08em
    T\kern-.1667em\lower.7ex\hbox{E}\kern-.125emX}}
    
\newtheorem{definition}{Definition}
\usepackage{xspace}
\newcommand*{\Scale}[2][4]{\scalebox{#1}{$#2$}}%
\newcommand{\fastsynth}{\textsf{fastsynth}\xspace}%
\newcommand{\Fastsynth}{\textsf{Fastsynth}\xspace}%
\usepackage{pgf}
\usepackage{tikz}
\usetikzlibrary{shapes.geometric,arrows,automata}
\usepackage{caption}
\usepackage{subcaption}
\usepackage{multirow}
\usepackage{pgfplots}
\usepackage{tabularx}
\usepackage{enumitem}

\algdef{SE}[SUBALG]{Indent}{EndIndent}{}{\algorithmicend\ }%
\algtext*{Indent}
\algtext*{EndIndent}
\usepackage{framed}

\newcommand*{\skipnumber}[2][1]{%
	{\renewcommand*{\alglinenumber}[1]{} #2}%
	\addtocounter{ALG@line}{-#1}}
\makeatother

\renewcommand{\Comment}[2][.5\linewidth]{%
	\leavevmode\hfill\makebox[#1][l]{$\triangleright$~#2}}

\begin{document}

\title{Learning Concise Models \\from Long Execution Traces}

\author{\IEEEauthorblockN{Natasha Yogananda Jeppu, Thomas Melham, Daniel Kroening}
	\IEEEauthorblockA{\textit{Department of Computer Science} \\
		\textit{University of Oxford, UK}\\
		natasha.yogananda.jeppu@cs.ox.ac.uk}
	\and
	\IEEEauthorblockN{John O'Leary}
	\IEEEauthorblockA{\textit{Intel Corporation}\\
		\textit{Portland, Oregon} \\
		john.w.oleary@intel.com}
}

\maketitle

\begin{abstract}
Abstract models of system-level behaviour have applications in design
exploration, analysis, testing and verification.  We describe a new
algorithm for automatically extracting useful models, as automata, from
execution traces of a HW/SW system driven by software exercising a use-case
of interest.  Our algorithm leverages modern \textit{program synthesis}
techniques to generate predicates on automaton edges, succinctly describing
system behaviour.  It employs trace segmentation to tackle complexity for
long traces.  We learn concise models capturing transaction-level,
system-wide behaviour—experimentally demonstrating the approach using traces
from a variety of sources, including the x86 QEMU virtual platform and
the Real-Time Linux kernel.
\end{abstract}


\section{Introduction}

In modern system design, hardware and software components are designed by
different teams---perhaps even in different companies---and integrated only
when first silicon or a stable hardware emulation model is available.  Early
on, it is hard to see how a given hardware IP will fare under a software
workload and how software behaviour is affected by hardware design
decisions.  Co-design of hardware and software to meet desired
specifications is therefore very challenging.

Emulation and virtual platforms provide a way to exercise software
components in an environment that simulates the eventual hardware behaviour. 
They can easily be instrumented to record execution traces for analysis, but
the traces generated are large and unstructured.  So they are difficult to
correlate with a high-level view of the system and its requirements. 
Concise, human-readable models that express high-level hardware-software
interactions can provide users with a better insight into the working of the
system.  This can, in turn, aid in design exploration, analysis, testing and
verification applications.

Several methods have been proposed to reverse-engineer automata that model
system behaviour from execution
traces~\cite{Biermann:1972:SFM:1638603.1638997, edsm, state_merge,
Heule2013, Walkinshaw2016}, but the labels on transition arcs are limited to
Boolean events that are explicit in the traces.  Realising abstract and
understandable models requires the user to know what abstract conditions are
significant in the system evolution, for example `the FIFO became more than
half-full', and to instrument the system to record such conditions. 
Extensions of these traditional
algorithms\cite{model_daikon,compute_walkinshaw} generate Extended
Finite-State Machines (EFSMs) with syntactically-expressed predicates on
transitions edges, but require a substantial number of trace samples from
simulation of the learned model for predicate inference.

In this paper we present a new algorithm for model learning that employs
program synthesis~\cite{gulwani2017program} to construct transition
predicates that are not explicit in the trace data.  Program synthesis has
high computational complexity, so our algorithm uses a trace segmentation
strategy to make it scalable to long execution traces.  In principle, the
algorithm should be applicable to trace data obtained from any source:
modelling, emulation, simulation, or the system itself.  But in this paper,
we focus our experiments on systems modelled by virtual platforms or
directly in software, and illustrate the motivation for our work in this
setting.  In Section~\ref{sec:results}, we show that our algorithm achieves
high fidelity to published data-sheet diagrams for high-level system
behaviour.

\medskip

\noindent \textbf{Contribution.} The primary contribution of this paper is a
new, scalable method for learning finite-state models from execution trace
data that produces abstract, concise models.  Our algorithm integrates a
SAT-based approach with program synthesis techniques.  It learns succinct
models from traces without the additional information that is typically
required by state-merging~\cite{Biermann:1972:SFM:1638603.1638997}.  The
resulting models also feature informative transition-edge predicates that
are not explicit in the trace.  We make an algorithmic improvement to model
learning by making it scale to long traces with a segmentation approach.

\section{Formal Model}

We suppose that we can collect execution traces of the system we are
interested in by observing a finite set of user-defined variables, $X=
\{x_{1}, \dots, x_{k}\}$, over some domain~$D$ (we simplify
presentation by assuming all variables have the same domain.) The set $X'=
\{x_{1}', \dots, x_{k}'\}$ contains corresponding primed variables,
also over domain~$D$.  A primed variable $x_i'$ represents an update to
the unprimed variable~$x_i$ at the end of a discrete step. 
The variables in $X$ could stand for concrete values directly observable in the
system or some function of such values, depending on the user's intent. 
A~\emph{valuation} $v: X \to D$ maps the variables in $X$ to values in~$D$. 
An \textit{observation} at time step $t$ is a valuation of the variables at
that time, and is denoted by $v_{t}$.  A \textit{trace} is a sequence of
observations over time; we write a trace $\sigma$ with $n$ observations as a
sequence of valuations $\sigma = v_{1}, v_{2}, \dots, v_{n}$.

Our aim is to construct an automaton from a trace to represent behaviour captured
by the trace.  The learned automaton is a Non-Deterministic Finite Automaton
(NFA).

\begin{definition}[Non-Deterministic Finite Automaton] \label{nfa_definition}
An NFA $\mathcal{M}\allowbreak=(\allowbreak\mathcal{Q},\allowbreak
q_0,\allowbreak\Sigma, \allowbreak{F}, \allowbreak\delta)$ is a state
machine where $\mathcal{Q}$ is a finite set of states, $q_0 \in
\mathcal{Q}$ is the initial state, $\Sigma$ is a finite alphabet,
${F}\subseteq\mathcal{Q}$ is the set of accepting states, and $\delta:
\mathcal{Q} \times \Sigma \rightarrow \mathcal{P(Q)}$ is the transition
relation.
\end{definition}

In our setting, all states of the automaton are accepting states, i.e., our
automaton rejects by running into a `dead end'.  A~symbol~$a$ of the alphabet~$\Sigma$
is a function $a : (X \cup X') \to D$, i.e., a pair of observations of
the system.  Let $\sigma = v_{1}, \dots, v_{n}$ be a trace of the system.
The symbol $a_{i}$ for $i = 1, \dots, n{-}1$ is
\begin{equation}
\begin{array}{l}
a_{i}(x) = v_{i}(x)\\
a_{i}(x') = v_{i+1}(x) 
\end{array}  
\end{equation}

\noindent The automaton accepts a word $w = a_{1}, \dots,
a_{p}$ over $\Sigma$ for $p < n$ if there exists a sequence of automaton states
$q_{1}, \dots, q_{p{+}1}$ such that
\begin{itemize}
\item $q_{1} = q_{0}$
\item $q_{i+1} \in \delta(q_{i},a_{i})$ for $i = 1, \dots, p$.
\end{itemize} 

\medskip

\section{Model Learning with Program Synthesis}\label{sec:algorithm}

Our model-learning algorithm is provided in Algorithm~\ref{algo}.
The algorithm fits into the following overall framework:
\begin{enumerate}[label=\textit{\Alph*.}]
\item \textit{Tracing infrastructure.} This records traces and performs any
required pre-processing.
\item \textit{Transition predicate synthesizer.} This generates transition
predicates from trace data using program synthesis.
\item \textit{Model construction algorithm.}  This iteratively constructs an
automaton from a sequence of predicates obtained from the previous step and
checks its compliance with the sequence input.
\end{enumerate} 

\noindent We describe these components in detail in the sections that follow.

%

\subsection{Tracing Setup}

We use implementations of the system of interest to obtain trace data.  For
most of our experiments, execution traces are produced simply by
instrumenting source code with print statements.  This provides flexibility
in getting the required information from the simulation runs.  Target
components of interest are identified and trace statements added at relevant
points in the source code based on the end goal for analysis.  Traces can
also be produced using any other means, for example inbuilt tracing or
logging frameworks.

\subsection{Transition Predicate Synthesis}

For observations that include non-Boolean variables, we need a way to
consolidate the information they represent into expressions that will serve
as transition predicates in our automaton.  We use a program synthesiser to
generate a state transition function $\mathit{next}(x)$ that provides the value
of the given variable $x \in X$ in the next state.  The method used is an
instance of \textit{synthesis from examples}~\cite{Gulwani2012SynthesisFE}. 
There are many algorithms that implement this synthesis technique.  We
discuss choices for the synthesis algorithm in
Section~\ref{sec:synthengines}.

Trace data are used to provide concrete samples for $\mathit{next}(x)$,
which in turn serve as constraints in the synthesis tool (line 3).  For
example, consider a system with a single variable $x_1$, and let $x_1{=}1$,
$x_1{=}2$, $x_1{=}3$, $x_1{=}4$ be a trace of that system.  The
corresponding examples for deriving $\mathit{next}(x_1)$ are
${\mathit{next}(1) = 2,\, \mathit{next}(2) = 3,\, \mathit{next}(3) = 4}$.
%
%
From these examples, the synthesis tool might generate $\mathit{next}(x_1) = x_1 + 1$, 
which we then use to model the behaviour of the variable $x_{1}$
in our NFA.  Consider another system with two variables $X =
\{x_{1},x_{2}\}$ and suppose the next-state value of $x_{1}$ depends on $x_{2}$:
\begin{equation}\label{formula:next2}
 x_{1}' =
\begin{cases}
      x_{1} + 1 & \text{ if $x_{2} = 0$}\\
      x_{1} - 1 & \text{ if $x_{2} = 1$}
 \end{cases}       
\end{equation}
One trace for this system might be this: $(x_1{=}1,\,x_2{=}0)$,
$(x_1{=}2,\,x_2{=}0)$, $(x_1{=}3,\,x_2{=}1)$,
$(x_1{=}2,\,\allowbreak x_2{=}0)$.
The corresponding examples for synthesis of a definition of $\mathit{next}(x_1,x_2)$ are
$\Scale[0.9]{\mathit{next}(1,0) = (2,0)}$,
$\Scale[0.9]{\mathit{next}(2,0) = (3,1)}$, and
$\Scale[0.9]{\mathit{next}(3,1) = (2,0)}$.

The synthesised function $\mathit{next}(x)$ is used to define a predicate
`$x'=\mathit{next}(x)$' that relates observations in the current and next
states, and will serve as a transition predicate in our learned automaton. 
To tackle the problem of synthesis complexity for long traces, a sequence
of these predicates is generated by feeding segments of the trace one after
the other into the predicate synthesis algorithm, using a sliding window of a suitable length depending on the application and the amount of data to be consolidated into predicates (lines 8--12).

\begin{algorithm}[t]
	\scriptsize
	\caption{Model Learning Algorithm}\label{algo}
	\begin{algorithmic}[1]
		\Procedure{GeneratePredicate}{$\sigma'$ : trace of length $w'$}
		\State $\sigma' = v_{i}, v_{i+1}, \dots, v_{i+w'-1}$
		\State $\mathit{next}(v_{k}) \gets v_{k+1}$, for $k = i,i+1,\dots,(i+w'-2)$
		\State Synthesize $\mathit{next}(x)$ 
		\State \textbf{return} `$(x' = \mathit{next}(x))$'
		\EndProcedure
		\medskip
		\Procedure{GenerateModel}{$\sigma$ : trace of length $n$}
		\State Divide $\sigma$ into $\{\sigma_{1}, \sigma_{2}, \dots, \sigma_{k} \}$, $k=(n+1-w')$
		\For{each $\sigma_{i}$}
		\State $p_{i} \gets \textsc{GeneratePredicate}(\sigma_{i})$
		\EndFor
		\State $P \gets p_{1}, p_{2}, \dots, p_{k}$\
		\State Target automaton $\mathcal{M}$ is represented as	\skipnumber[1]{\State \quad $\mathcal{M} = (q_{1},p'_{1},q'_{1}), (q_{2},p'_{2},q'_{2}), \dots, (q_{m},p'_{m},q'_{m})$} 
		\State $w \gets$ sliding window size
		\State Divide $P$ into $\{P_{1}, P_{2}, \dots, P_{k+1-w} \}$ where \skipnumber[1]{\State \quad $P_{i} = p_{i}, p_{i+1}, \dots, p_{i+w-1}$}
		\State $N \gets 1$\Comment{Number of automaton states}
		\State Constraint $c_0 = \exists i,j \in \{1,\ldots,m\} \ni (q_{i} = q_{j} \wedge p'_{i} = p'_{j} \wedge q_{i}' \neq q_{j}')$
		\State Set of blocking constraints $C \gets \{c_0\}$ 
		\State Generate the following C program:{\label{cbmc_gen_model}}
		
		\begin{leftbar}
			\State  assume $1 \leq q_{i},q_{i}' \leq N$, for $i=1,2,\dots ,m$ 
			\State $j \gets 0$ 
			\For{each $P_{i}$} 
			\For {$y = i$ {\normalfont\textbf{to}} $i+w-1$} 
			\State $p'_{j} \gets p_{y}$
			\State assume $q_{j+1} = q'_{j}$
			\State $j \gets (j+1)$
			\EndFor
			\EndFor
			\State assert $\underset{c \in C}{\bigvee} c$ 
		\end{leftbar}
		\State Run CBMC with the above program as input{ \label{assert}} 
		\If{assertion holds} \Comment{No candidate automaton found}
		
		\State $N \gets (N+1)$ 
		\State \textbf{go to} \ref{cbmc_gen_model}
		
		\Else \Comment{Candidate automaton $\mathcal{M}$ found}
		\State $\mathcal{M} \gets$ learned candidate automaton from counterexample
		\State $l \gets$ length of transition sequence for subsequence check
		\State $S_{l} \gets$ set of all transition sequences of length $l$ in $\mathcal{M}$
		\State $P_{l} \gets$ set of all subsequences of $P$ of length $l$
		\If{$S_{l} \nsubseteq P_{l}$} \Comment{Subsequence check failed}
		
		\State Encode sequences ($S_{l}-P_{l}$) as blocking constraints and add to $C$
		\State \textbf{go to} \ref{cbmc_gen_model}
		
		\Else \Comment{Subsequence check successful}
		
		\State \textbf{return} $\mathcal{M}$
		\EndIf
		\EndIf
		\EndProcedure
	\end{algorithmic}
\end{algorithm}

\begin{figure*}[h]
	\begin{subfigure}[b]{\columnwidth}
		\centering
		\includegraphics[width=1.1\textwidth]{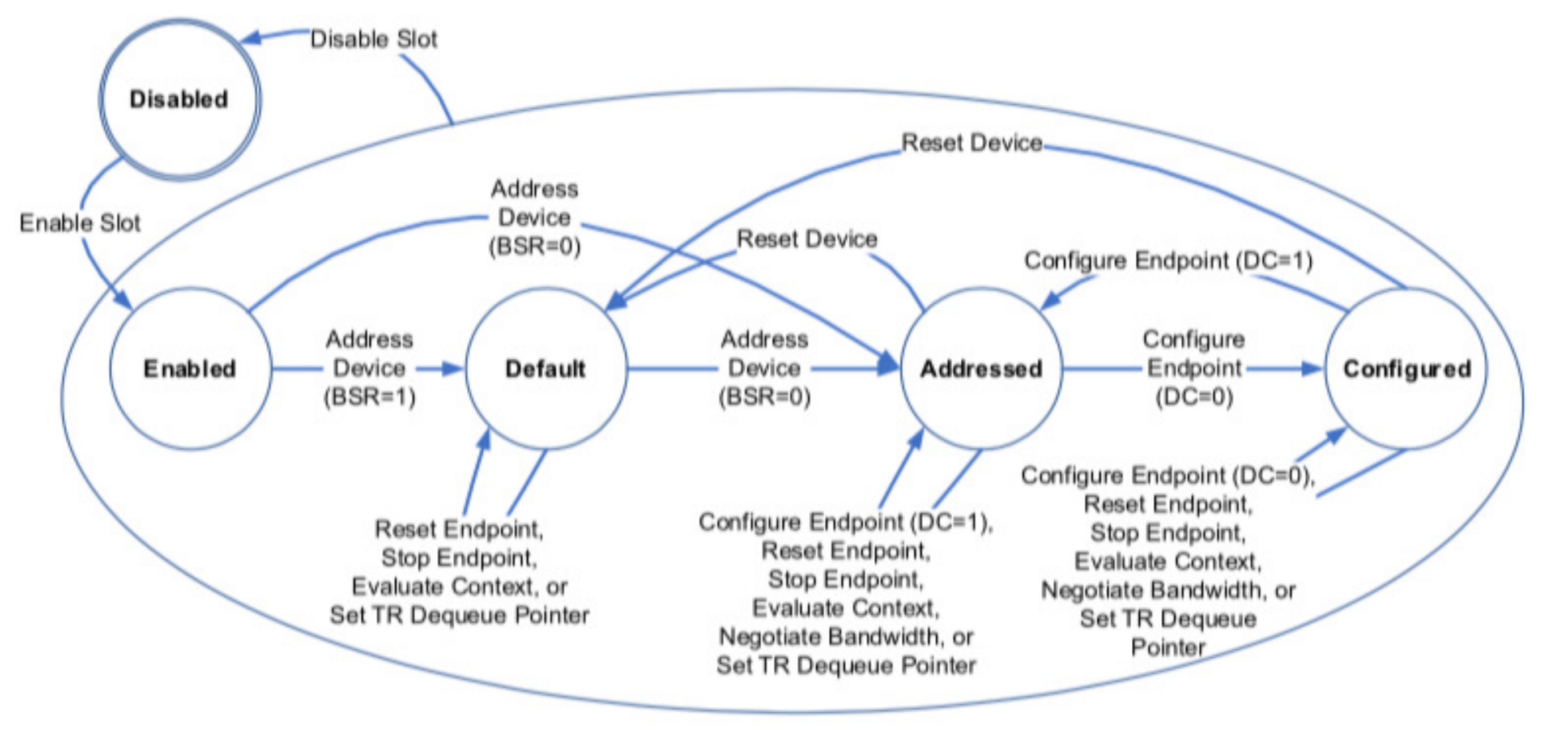}
		\caption{}
		\label{fig:datasheet_model}
	\end{subfigure}
	\begin{subfigure}[b]{\columnwidth}
		\centering
		\begin{tikzpicture}[->,>=stealth',shorten >=1pt,auto,node distance=2.5cm, thick]
		\tikzstyle{every state}=[fill=white,draw=black,text=black]
		\node[initial,state] (A)                    {$q_1$};
		\node[state]         (B) [below of=A] {$q_2$};
		\node[state]         (D) [right of=B,xshift=1cm] {$q_3$};
		\node[state]         (C) [right of=D,xshift=1cm] {$q_4$};
		
		\path (A) edge [bend right]      node[anchor=west]{\scriptsize CR\_ENABLE\_SLOT} (B)
		(C) edge [bend right]            node[anchor=south,above = 2mm]{\scriptsize CR\_DISABLE\_SLOT} (A)
		(C) edge [bend right]                node [anchor=south]{\scriptsize CR\_RESET\_DEVICE} (B)
		(B) edge [left]            node[align=center,anchor=north]{\scriptsize CR\_ADDR\_DEV \\ \scriptsize (BSR=0)} (D)
		(D) edge [bend left]            node[anchor=north,below=2mm]{\scriptsize CR\_CONFIG\_END} (C)
		(C) edge [bend left]            node[anchor=north]{\scriptsize CR\_STOP\_END} (D)
		(C) edge [loop below]            node[anchor=north,xshift=-7mm]{\scriptsize CR\_CONFIG\_END} (C);
		\end{tikzpicture}
		\caption{}
		\label{fig:model_gen}
	\end{subfigure}
	\caption{USB Slot state machine provided in (a) Intel datasheet~\cite{intel_datasheet} and (b) model learnt by our framework.}
\end{figure*}
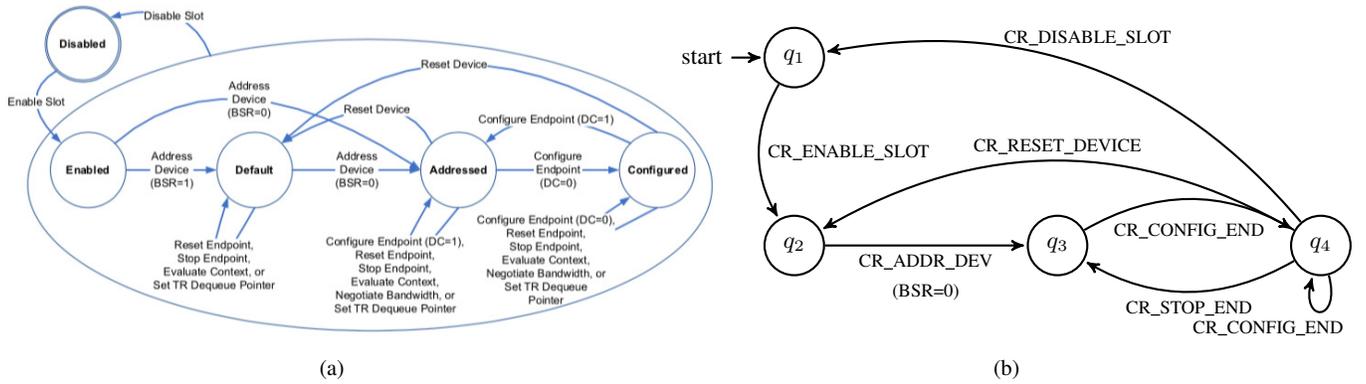

\subsection{Model Construction}

Our model construction algorithm takes as input a sequence of predicates $P
= p_{1}, \dots, p_{k}$ of the form just described (line 12).  Each predicate
is represented by an expression (a syntax tree) over variables in $(X \cup
X')$.  The automaton $\mathcal{M}$ to be constructed is represented as an
array, each element of which encodes a transition.  The $i$-th element in
the array is a triple comprising the following symbolic variables: a state
variable $q_{i}$ for the state from which the transition occurs, a variable
$p'_{i}$ for the corresponding transition predicate and a next state
variable $q'_{i}$ for the state the system moves to (line 13).  The sequence
of predicates is divided into segments using a sliding window $w$, a
parameter that can be tuned, and unique segments are processed further (lines
14--15).  These predicate segments are later used to encode transition sequences
in the automaton.  The parameter $w$ determines the input size, and
consequently the algorithm runtime.  Choosing $w=1$ will not capture any
sequential behaviour but only ensures that all trace events appear in the
automaton.  For model learning, we would like to choose a value for $w$ that
results in a small input size but is not trivial $(w = 1)$.  For our
experiments we performed multiple runs of the algorithm, randomly selecting
a different value for $w$ between $2 \le w \le |P|$ for each run, and
obtained the same automaton in all scenarios.  The strategy we adopt for our
experiments is to fix window length $w=3$, which is small to ensure quick
results, and yet captures interesting trace patterns.  The result of
segmentation is the \emph{set} of all unique subsequences of $P$ of
length~$w$.  Segmentation significantly improves runtime, especially for
long traces, by leveraging the presence of repeating patterns in the trace. 
A detailed discussion of the need for segmentation and its effect on
scalability is given in Section~\ref{sec:seg}.

To construct the model, we search systematically for an $N$-state NFA whose
behaviours include all the unique segments previously identified and has at
most one transition from any state labelled with any given predicate. The latter is encoded as the negation of the desired property (line 17) and added to a set of blocking constraints that we maintain to restrict the automaton throughout the construction process (lines 17--18). For a
given $N$, we hypothesize no such automaton exists and use a model checker
to check the hypothesis.  This is done by first encoding the hypothesis as a C program (lines 19--29).  For the check we use the C Bounded Model
Checker~(CBMC)~\cite{ckl2004}.  We fix
the number of automaton states by restricting the state variables of
$\mathcal{M}$ to take values between 1 and $N$ (line 20).  Lines 21--28
ensure that the automaton always includes the corresponding transition
predicates in the sequence they appear in the segments of~$P$. We assert that at least one of the blocking constraints is true (line 29). The program, along with this assertion, is then fed to CBMC (line 30). If the assertion holds (line 31), it indicates that for all assignment of values in the range $1$ to $N$ to
state variables $q_{i}$ and $q_{i}'$ of $\mathcal{M}$, at least one of the blocking constraints on the automaton hold. This implies that there is no $N$-state automaton
that meets our specifications; in this case we increment $N$ and repeat the
search (lines 31--33).  We begin model construction with $N=1$ and increase
the number of states by $1$ if such an automaton cannot be learned.  This
ensures that we learn the smallest automaton that contains all subsequences
of $P$ of length~$w$.

A counterexample to the assertion is an assignment of values to state
variables of $\mathcal{M}$ that encodes an $N$-state automaton that contains all subsequences
of $P$ of length~$w$ and also, does not satisfy any of the blocking constraints (line 34).  Once CBMC has constructed a candidate
model, we check its compliance with $P$ by looking for \emph{invalid}
transition sequences in $\mathcal{M}$ (lines 35--39).  A transition sequence
is said to be invalid if it is not a subsequence of~$P$.  We check if all
transition sequences in the model of a given length $l$ are subsequences
of~$P$ (line 39).  The parameter $l$ can be tuned to change the degree of
generalisation.  However, a higher value for $l$ implies tighter constraints
on the model moving towards a more exact representation.  It is known that
learning exact automata from trace data is
NP-complete~\cite{Gold1978ComplexityOA}.  Hence, we have used $l=2$ to
ensure that it is not too complex for the model checker to solve but at the
same time does not over-generalise to fit the trace.  We~encode invalid
sequences as additional blocking constraints on the automaton and repeat the search
(lines 39--41).  This refinement loop gleans further information from
sequence $P$.  The algorithm returns the $N$-state automaton
$\mathcal{M}$ if such a candidate automaton is found and the compliance check is
successful (lines 42--43).

\section{Benchmarks}\label{sec:results}

We demonstrate our model learning approach from execution traces for six
examples.  We compare our approach against the traditional state-merge
algorithm and provide experimental evidence to demonstrate scalability of
our algorithm.  Four of the six benchmarks use the QEMU virtual platform to
emulate an x86 system, including the hardware components.  The virtual
platform runs a full CentOS Linux distribution and is given an application
to exercise system behaviours of interest.  The other two benchmarks are
artificial and enable us to benchmark particular aspects of our method.

\medskip

\noindent \textbf{USB xHCI Slot State Machine}
The Extensible Host Controller Interface (xHCI) specifies a controller's
register-level operations for USB 2.0 and above.  In this example we look
specifically at slot-level operations that take place when we access a
virtual USB storage device as implemented in QEMU x86 platform emulation. 
The framework learns the automaton given in Fig.~\ref{fig:model_gen}
resembling the Intel datasheet diagram in Fig.~\ref{fig:datasheet_model}.

It is worth noting that the model learning algorithm is able to generate an
accurate representation of system behaviours that are exercised under a
given application load.  Some transitions in Fig.~\ref{fig:datasheet_model}
do not appear in the learned model because either QEMU does not implement
those scenarios or that the application load does not drive the system into
those scenarios.  The models generated thus provide valuable coverage
information.

\medskip

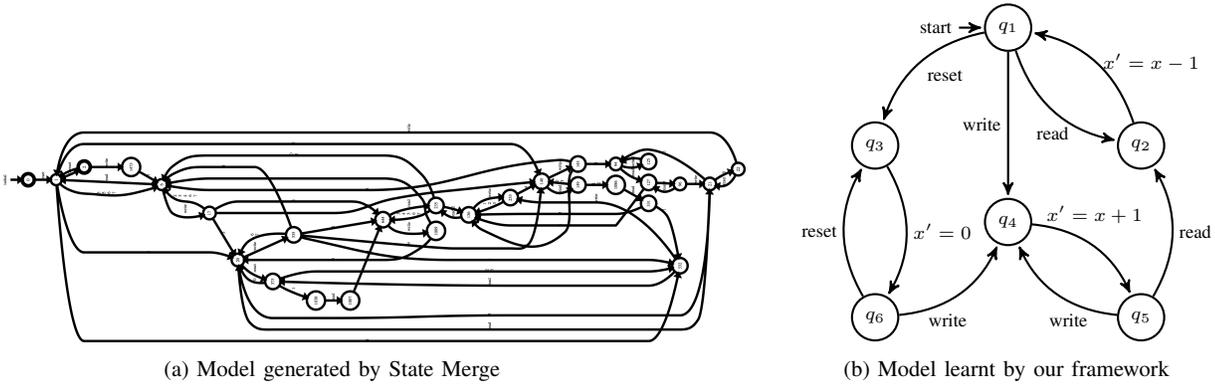
\begin{figure*}[h]
	\begin{subfigure}[b]{\columnwidth}
		\centering
		\includegraphics[width=1.12\textwidth]{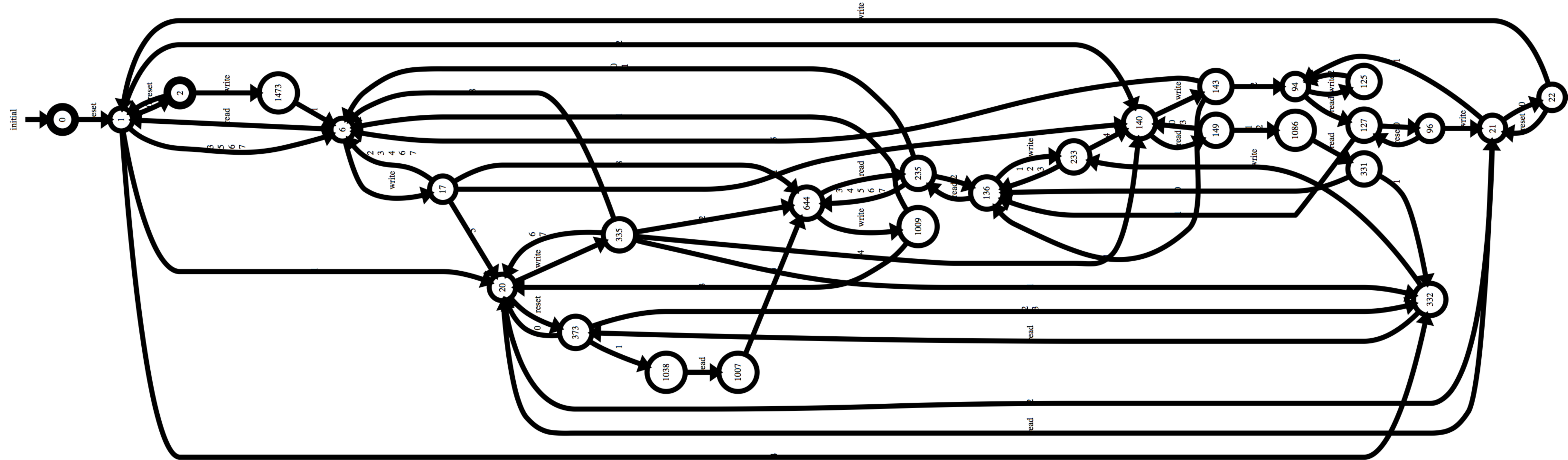}
		
		\caption{Model generated by State Merge}
		\label{fig:uart_statemerge}
	\end{subfigure}
	\begin{subfigure}[b]{\columnwidth}
		\centering
		\scriptsize
		\begin{tikzpicture}[->,>=stealth',shorten >=1pt,auto,node distance=2.5cm,
		thick]
		\tikzstyle{every state}=[fill=white,draw=black,text=black]
		\node[initial,state] (A)                    {$q_1$};
		\node[state]                   (B) [below right of=A, yshift=0.2cm] {$q_2$};
		\node[state]                   (E) [below left of=A, yshift=0.2cm]  {$q_3$};
		\node[state]                   (F) [below left of=B,  yshift=0.75cm] {$q_4$};
		\node[state]                   (C) [below right of=F, yshift=0.5cm]  {$q_5$};
		\node[state]                   (D) [below left of=F,  yshift=0.5cm]  {$q_6$};
		\path (A) edge [bend right]             node [anchor=east,yshift=-3mm,xshift=3mm]{read} (B)
		(B) edge [bend right]       node [anchor = west]{$x' = x-1$} (A)
		(C) edge [bend right]      node [anchor=west]{read} (B)
		(C) edge [bend left]       node [anchor=east,yshift=-3.5mm,xshift=5mm]{write} (F)
		(F) edge [bend left]       node [yshift=2mm,xshift=-7mm]{$x' = x+1$} (C)
		(D) edge [bend right]      node [anchor=west,yshift=-3.5mm,xshift=-5mm]{write} (F)
		(D) edge [bend left]       node {reset} (E)
		(A) edge [left]            node {write} (F)
		(A) edge [bend right]      node {reset} (E)
		(E) edge [bend left]       node [anchor=west]{$x' = 0$} (D);
		\end{tikzpicture}
		\caption{Model learnt by our framework}
		\label{fig:uart}
	\end{subfigure}
	\caption{QEMU Serial I/O Port}
\end{figure*}

	\begin{figure}[h]
	\centering
	\scriptsize
	\begin{tikzpicture}[->,>=stealth',shorten >=1pt,auto,node distance=2cm,
	thick]
	\tikzstyle{every state}=[fill=white,draw=black,text=black]
	\node[initial,state] (A)                    {$q_1$};
	\node[state]         (B) [right of=A] {$q_2$};
	\node[state]         (C) [above right of=B] {$q_3$};
	\node[state]         (D) [below right of=B] {$q_4$};
	\node[state]         (E) [below right of=C] {$q_5$};
	\node[state]         (F) [below right of=D,yshift=-3mm] {$q_6$};
	\node[state]         (G) [below left of=D,yshift=-3mm] {$q_7$};
	\path (A) edge [right]             node [anchor=south]{\scriptsize xhci\_write} (B)
	(B) edge [right]            node [anchor=east]{\scriptsize ErTransfer} (C)
	(B) edge [right]            node [anchor=west,above=4mm,right=-2mm] {\scriptsize ErCC, ErPSC} (D)
	(C) edge [right]            node [anchor=west]{\scriptsize CCSuccess} (E)
	(D) edge [right]            node [anchor=east] [anchor=east,above=1mm,left=-1mm]{\scriptsize CCSuccess} (E)
	(E) edge [bend left]            node [anchor=west] {\scriptsize xhci\_ring\_fetch} (F)
	(F) edge [bend left]            node [anchor=east,below=5mm,left=0.05mm] {\scriptsize TRData, TRSetup} (E)
	(F) edge [bend left]            node  [anchor=north]{\scriptsize TRBReserved} (G)
	(G) edge [bend left]            node  [anchor=south]{\scriptsize xhci\_ring\_fetch} (F)
	(G) edge [left]            node [anchor=east,above=6mm,left=0.1mm]  {\scriptsize xhci\_write} (B)
	(F) edge [bend left]            node [align=center,above=10mm,left=13mm] {\scriptsize CrAD, CrCE, CrES\\\scriptsize TRNormal, TRStatus} (A);
	\end{tikzpicture}

	\caption{Model of USB interface learnt by our framework.}
	\label{fig:usb2}
\end{figure}
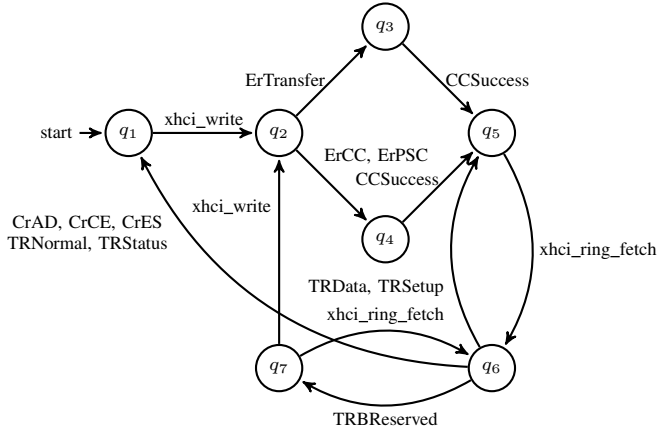
\begin{figure}[h]
	\centering
	\begin{tikzpicture}[->,>=stealth',shorten >=1pt,auto,node distance=3cm,
	thick]
	\tikzstyle{every state}=[fill=white,draw=black,text=black]
	\node[initial,state] (A)                    {$q_1$};
	\node[state]         (B) [below right of=A] {$q_2$};
	\node[state]         (C) [below left of=A] {$q_3$};
	\path (A) edge [loop above]             node {$op' = op + ip$} (A)
	(A) edge [bend left]             node [anchor=west,xshift=2mm] {$reset$} (B)
	(B) edge [bend left]            node [anchor=east,yshift=-7mm,xshift=9mm]{$op' = 0$} (A)
	(A) edge [bend right]            node [anchor=east,xshift=-6mm] {$(op=5 \allowbreak \wedge
		\allowbreak ip=1) \allowbreak \vee \allowbreak$} (C)
	(A) edge [bend right]            node [anchor=east,yshift=-4mm,xshift = -4mm] {$(op=-5 \allowbreak \wedge \allowbreak ip=-1)$} (C)
	(C) edge [bend right]            node [anchor=west,yshift=-5mm,xshift = -5mm] {$op'=op$} (A);
	\end{tikzpicture}
	\caption{Model of an anti-windup integrator learnt by our framework.}
	\label{fig:integrator}
\end{figure}
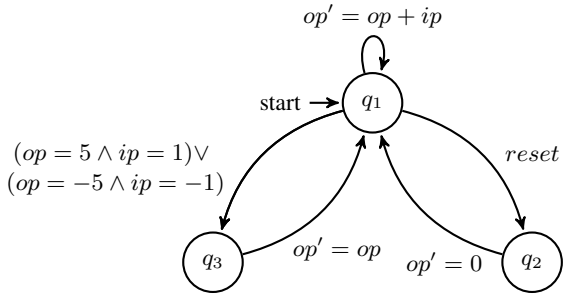 

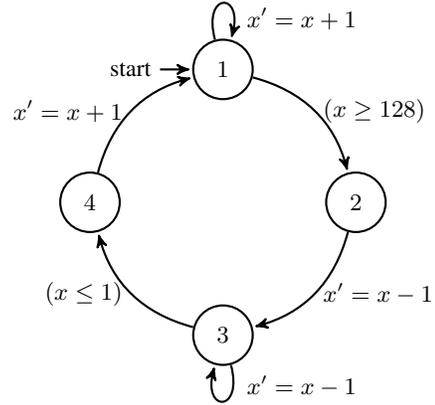
\begin{figure}[h]
	\centering
	\begin{tikzpicture}[->,>=stealth',shorten >=1pt,auto,node distance=2.5cm,
	thick]
	\tikzstyle{every state}=[fill=white,draw=black,text=black]
	\node[initial,state] (A)                    {$1$};
	\node[state]         (B) [below right of=A] {$2$};
	\node[state]         (C) [below left of=B] {$3$};
	\node[state]         (D) [below left of=A] {$4$};
	\path (A) edge [bend left]  node[anchor=west]{$(x\ge 128)$} (B)
	(A) edge [loop above]       node[anchor=west,right=2mm,yshift=-2mm]{$x' = x+1$} (A)
	(B) edge [bend left]        node[anchor=west]{$x' = x-1$} (C)
	(C) edge [bend left]        node[anchor=east]{$(x\le 1)$} (D)
	(C) edge [loop below]       node[anchor=west,right=2mm,yshift=2mm]{$x' = x-1$} (C)
	(D) edge [bend left]        node[anchor=east]{$x' = x+1$} (A);
	\end{tikzpicture}
	
	\caption{Model of a counter with threshold 128 learnt by our framework.}
	\label{fig:count128}
\end{figure}

\begin{figure*}[h]
	\centering
	\scriptsize
	\begin{tikzpicture}[->,>=stealth',shorten >=1pt,auto,node distance=1.8cm,
	thick]
	\tikzstyle{every state}=[fill=white,draw=black,text=black]
	\node[initial,state] (A)                    {$q_1$};
	\node[state]                   (C) [below right of=A,xshift=1cm] {$q_3$};
	\node[state]                   (B) [below left of=C,xshift=-1cm] {$q_2$};
	\node[state]                   (E) [right of=C,xshift=1cm]  {$q_4$};
	\node[state]                   (D) [above of=E,  yshift=-0.6cm] {$q_5$};
	\node[state]                   (F) [right of=E,xshift=3cm]  {$q_7$};
	\node[state]                   (G) [right of=D,xshift=3cm]  {$q_8$};
	\node[state]                   (H) [below of= F,yshift=0.75cm]  {$q_6$};
	\path (A) edge [bend left=20]             node [anchor=south]{sched\_switch\_in} (G)
	(A) edge [bend right]             node [anchor=east]{sched\_waking} (B)
	(B) edge [bend right]             node [anchor=west,yshift=0.4cm]{sched\_waking} (A)
	(B) edge [bend right=20]             node [anchor=north]{sched\_switch\_in} (H)
	(C) edge [right]             node [anchor=north,sloped]{sched\_waking} (B)
	(D) edge [right]             node [anchor=east,yshift=0.25cm,xshift=0.5cm]{sched\_switch\_suspend} (C)
	(D) edge [right]             node [anchor=south,sloped]{sched\_switch\_preempt} (A)
	(E) edge [right]             node [anchor=east,yshift=-0.15cm]{sched\_entry} (D)
	(B) edge [bend right=20]             node [anchor=west,xshift = 0.5cm]{set\_state\_sleepable,} (E)
	(B) edge [bend right=20]             node [anchor=west,yshift=-0.2cm]{set\_need\_resched} (E)
	(G) edge [bend left]             node [anchor=west]{set\_state\_sleepable} (F)
	(F) edge [bend left]             node [anchor=east]{sched\_waking} (G)
	(F) edge [bend left]             node [anchor=west]{set\_state\_runnable} (H)
	(H) edge [bend left]             node [anchor=east]{set\_state\_sleepable} (F)
	(F) edge [left]             node [anchor=west,yshift=0.7cm,xshift=-2cm]{sched\_entry} (D)
	(F) edge [left]             node [anchor=north]{set\_need\_resched} (E)
	(G) edge [left]             node [anchor=east,yshift=0.6cm,xshift=2cm]{set\_need\_resched} (E);
	\end{tikzpicture}
	\caption{Model of RT-Linux Kernel Thread Scheduling learnt by our framework}
	\label{fig:linux}
\end{figure*}
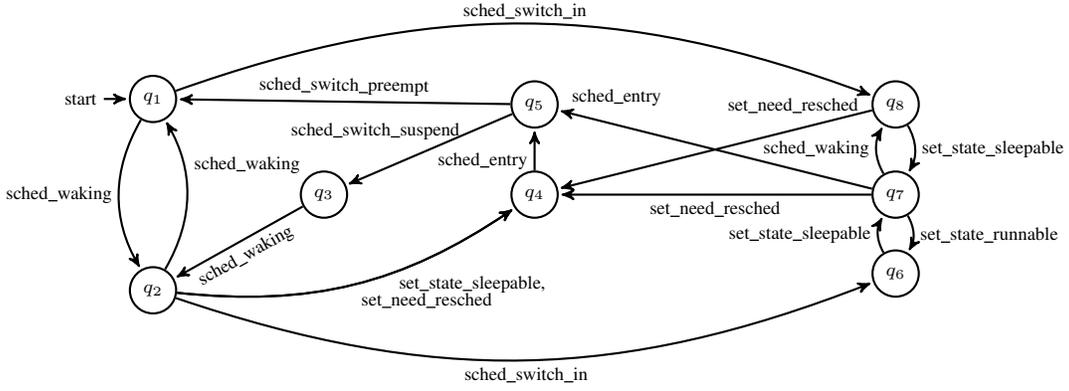

\noindent \textbf{QEMU USB Interface Emulation}
We use the same setup as above but record all interface events that
take place when a virtual USB storage device is attached to the virtual
platform.  The resulting trace records the series of ring fetch and ring
write operations between the command ring and event ring of the xHCI
protocol. Our algorithm learned a concise 7 state automaton, while the
smallest model generated by state-merge had 91 states
(Table~\ref{tab:state_merge}).  \medskip

\noindent \textbf{QEMU Serial I/O Port}
We used QEMU's x86 emulation of a serial I/O port and recorded changes in
queue length.  The trace contains (numerical) queue length data along with
(Boolean) read, reset and write events on the queue.  

Our algorithm was able
to generate expressions for transition predicates as given in
Fig.~\ref{fig:uart}.  We were unable to take the queue to its full capacity
due to very quick read-writes and frequent resets.

%

\medskip 

\noindent \textbf{Counter}
We trace a simple program that counts from $1$ up to a threshold~$T$, which
we set to $128$; after it reaches the threshold it counts back down to $1$. 
This is repeated $N$ times.  We observe the value of the counter. The
synthesis component of our algorithm automatically generates the expected
transition predicates $(x' = x+1)$, $(x \geq 128)$, $(x \leq 1)$ and $(x' =
x-1)$ using just the values from the trace. 

This benchmark is particularly interesting as it has constants, like the
value of $T$, in the predicates.  The ability to automatically produce these
constants depends on the synthesis tool used and the approach to synthesis. 
We discuss this in detail in  Section~\ref{sec:synthengines}.

\medskip

\noindent \textbf{Integrator}
Control applications frequently track an integral of an input signal.  We
implement an anti-windup integrator where the computed output $\mathit{op}$
is saturated at predefined thresholds, $5$~and~$-5$.  The input
$\mathit{ip}$ is restricted to take values $\{1,0,-1\}$.  The trace contains
valuations of $(\mathit{ip},\mathit{op})$ pairs at discrete time steps.  

The algorithm generates complex transition predicates $(op'=op+ip)$,
$(op'=op)$ and $(op=5 \allowbreak \wedge \allowbreak ip=1) \vee (op=-5
\allowbreak \wedge \allowbreak ip=-1)$.  The transitions on $(op'=op+ip)$
encode integrator behaviour outside saturation.  In the learned model,
transitions on $(op=5 \allowbreak \wedge \allowbreak ip=1) \vee (op=-5
\allowbreak \wedge \allowbreak ip=-1)$ are always followed by transitions on
$(op'=op)$; hence, accurately capturing behaviour at saturation.

\medskip

\noindent \textbf{Real Time Linux Kernel }
We generated an automaton describing the behaviour of threads in the Linux
PREEMPT\_RT kernel on a single core system.  This is motivated by work
in~\cite{10.1007/978-3-030-30446-1_17,linux} where hand-drawn models of the
kernel are used as monitors for runtime kernel verification.  For this
example, we used the built-in Linux tracing infrastructure \emph{ftrace} to
trace scheduler-related calls made by the thread under analysis, as
described in~\cite{linux}.  We used the Linux PREEMPT\_RT kernel version
5.0.7-rt5 on a single core QEMU emulated x86 machine for our
experiments.

Initial attempts at modelling thread behaviour with our algorithm, using the
\emph{pi\_stress} tests from the \emph{rt-tests} suite as system load,
revealed that some states in the hand-drawn model provided in~\cite{linux}
are not covered by the given load.  On running an additional kernel module
to cover these corner cases, we obtain the automaton in Fig~\ref{fig:linux}. 
This experiment provides evidence in support of potentially using the models
learned by our algorithm for functional test coverage analysis.

\section{The Benefit of Trace Segmentation}
\label{sec:seg}

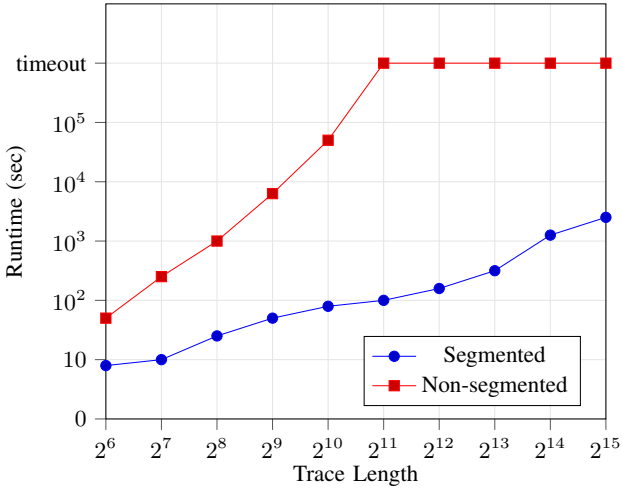
\begin{figure}[t]
	\centering
	\begin{tikzpicture}[scale=0.97]
	
	\begin{axis}[
	axis background/.style={fill=white},
	axis line style={black},
	tick align=outside,
	tick pos=left,
	x grid style={white!89.80392156862746!black},
	xlabel={Trace Length},
	xmajorgrids,
	xmin=1, xmax=10,
	xtick style={color=white!33.33333333333333!black},
	y grid style={white!89.80392156862746!black},
	ylabel={Runtime (sec)},
	ymajorgrids,
	ymin=0, ymax=7,
	ytick style={color=white!33.33333333333333!black},
	legend style={at={(0.95,0.2)},draw=black, fill=white},
	xtick = {1,2,3,4,5,6,7,8,9,10},
	ytick = {0,1,2,3,4,5,6},
	yticklabels = {$0$,$10$,$10^2$,$10^3$,$10^4$,$10^5$,timeout},
	xticklabels = {$2^6$,$2^7$,$2^8$,$2^9$,$2^{10}$,$2^{11}$,$2^{12}$,$2^{13}$,$2^{14}$,$2^{15}$}
	]
	
	\addplot coordinates {(1,0.9) (2,1) (3,1.4) (4,1.7) (5,1.9) (6,2) (7,2.2) (8,2.5) (9,3.1) (10,3.4)};
	\addplot coordinates {(1,1.7) (2,2.4) (3,3) (4,3.8) (5,4.7) (6,6) (7,6) (8,6) (9,6) (10,6)};
	\legend{Segmented, Non-segmented}
	\end{axis}
	
	\end{tikzpicture}
	\caption{Graph plot (log--log plot) comparing runtime for segmented and non-segmented trace input for the integrator example.}
	\label{fig:segment}
\end{figure}

To learn models from execution traces we require efficient and scalable
mechanisms for mining useful information from large amounts of trace data. 
More often than not, execution traces of a system contain recurring
patterns, which we exploit to speed up model learning.  To evaluate the
benefit of our segmentation technique, we give the results of a runtime
comparison of the algorithm for segmented and non segmented trace inputs for
all six examples (Table~\ref{tab:segment}).  We observe that the
segmentation enables our algorithm to scale: Fig.~\ref{fig:segment} is a
plot of the runtime against trace length for exponentially increasing trace
lengths for the integrator example.  Segmentation breaks down a large
problem into multiple small instances that have manageable runtime.  We
leverage the presence of trace patterns to significantly reduce execution
time as it is sufficient to process repeating segments once.

\begin{table}[t]
	\centering
	\scriptsize
	\begin{tabular}{|l|c|c|c|c|}
		\hline
		Example         & N & Trace Length    & Full Trace (s)        & \begin{tabular}[c]{@{}c@{}}Segmented\\ Trace (s)\end{tabular} \\ \hline
		USB Slot        & 4 & \phantom{000}39 & \phantom{00000}14.1     & \phantom{000}\textbf{9}\phantom{.0}  \\ \hline
		USB Attach      & 7 & \phantom{00}259 & \phantom{000}2249.5    &            \phantom{0}\textbf{915.4} \\ \hline
		Counter         & 4 & \phantom{00}447 & \phantom{0000}249.1    & \phantom{00}\textbf{95.9}  \\ \hline
		Serial I/O Port & 6 & \phantom{0}2076 & \phantom{00}23590.5 &            \phantom{00}\textbf{60.2} \\ \hline
		Linux Kernel             & 8 & 20165 &\textgreater 16 hours    & \phantom{0}\textbf{516.3}  \\ \hline
		Integrator      & 3 & 32768           & \textgreater 16 hours &            \textbf{3495.6} \\ \hline
	\end{tabular}
	\caption{Runtime comparison for segmented and non-segmented trace input. For a fair comparison, we begin learning with number of states equal to $N$.}
	\label{tab:segment}
\end{table}

\section{Comparison with State Merge Algorithms}

State merge algorithms are the established approach to model generation from
traces.  Traces are first converted into a Prefix Tree Acceptor (PTA). 
Model inference techniques are used to identify pairs of equivalent states
to be merged in the hypothesis model.  One of the most popular and accurate
inference techniques is Evidence-Driven State Merging (EDSM)~\cite{edsm}. 
The MINT (Model Inference Technique) tool~\cite{mint,Walkinshaw2016},
implements a variant of EDSM using data classifiers to classify trace events
based on next event.  States for which the classifier predicts the same next
event are merged.  It also provides support for the traditional $k$Tails
approach to state merge~\cite{Biermann:1972:SFM:1638603.1638997}.

A runtime comparison of MINT against our approach reveals that the
state-merge algorithm is significantly faster (Table~\ref{tab:state_merge})
but generates large automata that are difficult to comprehend, as shown
in~Fig.~\ref{fig:uart_statemerge}.  By contrast, our framework learns models
that are much more succinct (Fig.~\ref{fig:uart}) and capture system
behaviour accurately.  The MINT tool was unable to produce models from long
traces of length \textgreater 20,000 for the Linux kernel and integrator
examples whereas our approach successfully produced concise automata in both
cases.

\begin{table}[t]
\centering
\scriptsize
\begin{tabular}{|l|c|c|c|c|c|}
\hline
\multirow{2}{*}{\raisebox{-0.3cm}{Example}} & \multirow{2}{*}{\raisebox{-0.3cm}{\begin{tabular}[c]{@{}c@{}}Trace \\ Length\end{tabular}}} & \multicolumn{2}{c|}{Runtime (s)}                                    & \multicolumn{2}{c|}{Number of States}                               \\ \cline{3-6} 
                         &                  & \begin{tabular}[c]{@{}c@{}}State \\ Merge\end{tabular} & \begin{tabular}[c]{@{}c@{}}Model\\ Learning\end{tabular}  & \begin{tabular}[c]{@{}c@{}}State \\ Merge\end{tabular} & \begin{tabular}[c]{@{}c@{}}Model \\Learning\end{tabular} \\ \hline
USB Slot                 & \phantom{000}39  & \phantom{0}\textbf{8.7}                                & \phantom{00}14.5          & \phantom{00}6      & \textbf{4} \\ \hline
USB Attach               & \phantom{00}259  & \textbf{35.1}                                          & 3615.1                    & \phantom{0}91      & \textbf{7} \\ \hline
Counter                  & \phantom{00}447  & \textbf{12.1}                                          & \phantom{00}98.6          & 377                & \textbf{4} \\ \hline
Serial I/O Port          & \phantom{0}2076  & \textbf{28.6}                                          & \phantom{0}137.4          & \phantom{0}28      & \textbf{6} \\ \hline
Linux Kernel                      & 20165  &       $\approx$ 5 h                                 &    \textbf{4173.6}       & no model     &  \textbf{8}\\ \hline
Integrator               & 32768            & $\approx$ 5 h                                          & \textbf{3497.2} & no model           & \textbf{3} \\ \hline
\end{tabular}
\caption{Runtime analysis of state-merge vs.~model learning.}
\label{tab:state_merge}
\end{table}

\section{Discussion of Program Synthesis Engines}
\label{sec:synthengines}

Virtually all tools that perform program synthesis implement a form of
Counter Example Guided Inductive Synthesis
(CEGIS)~\cite{gulwani2017program}.  The program that is generated is usually
required to conform to a grammar, which is given as part of the problem
description.  Tools that require this grammar implement Syntax-Guided
Synthesis (SyGuS)~\cite{sygus}.  We experimented with two program synthesis
tools for generating transition predicates for the automaton: CVC4 version
1.6~\cite{cvc4, DBLP:conf/cav/BarrettCDHJKRT11}, which by default employs
SyGuS, and \fastsynth, which is based on the work done in~\cite{cegisT}.

The SyGuS-based approach requires a grammar for the program.  The key effort
is to determine the constants that are required; they have to be adjusted
manually for every model.  \Fastsynth also implements CEGIS but does not
rely on a user-specified grammar to restrict the search space.  \Fastsynth
ignores any grammar given as part of the problem and produces the smallest
function that satisfies the constraints.  Any constants that may be required
are generated automatically.

CVC4 also implements an alternative algorithm that does not require syntax guidance;
however, that produces trivial solutions.  For example, given the trace
sequence $1,\,2,\,4,\,8$, CVC4 generates
$ \Scale[0.9]{\textit{\textrm{(ite (= x 4) 8 (ite (not (= x 2)) 2 4))}}}$
whereas \fastsynth produces the expression $\textit{\textrm{x + x}}$, which
is a better fit for our problem.  The type of transition predicates
synthesised depends on the ability of the program synthesis tool and the
approach to synthesis.  A~suitable synthesis tool can be chosen based on the
target models we wish to obtain and the application domain.

\section{Related Work}\label{sec:related}

Manually creating abstract system models is time-consuming and error-prone,
and this has prompted numerous research efforts aimed at automated model
learning.  The most common approach to automatically reverse engineer models
from execution traces is state
merging~\cite{Biermann:1972:SFM:1638603.1638997}.  The process involves
converting traces into prefix tree acceptors (PTA), and then applying model
inference techniques to determine which states are to be merged.
%
In the traditional $k$Tails approach two states in the PTA are merged if they are \emph{k-equivalent}.
The parameter $k$ is used to change the degree to which the model generalises. A variant of the algorithm~\cite{Walkinshaw2016} additionally uses data classifiers to determine state equivalence. 

Conventional automata learning approaches are partial because they fail to
model how system variables change during execution.  An extension of the
state merge algorithm~\cite{compute_walkinshaw} generates ``computational"
state machines.  Data update functions over transitions are automatically
generated using genetic programming.
%
%
This method requires additional trace data, over and above the trace data
used to generate the initial model, for transition predicate inference.  The
GK-Tails~\cite{model_daikon} algorithm integrates Daikon~\cite{daikon} with
the $k$Tails approach to derive transition predicates for Extended Finite
State Machines (EFSM) that represent software behaviour.  The type of
expressions generated is however restricted.
%
%
$k$Tails based algorithms use instances of only positive behaviour and hence
run the risk of over-generalising~\cite{Gold1978ComplexityOA}.

A popular model inference algorithm, Evidence-Driven State Merge
(EDSM)~\cite{edsm}, overcomes the problem of over-generalisation by using
both positive and negative instances of behaviour to determine equivalence
of states to be merged based on statistical evidence.
%
In an extension of this work~\cite{Heule2013}, finite automata inference is mapped to a graph-colouring
problem based on the red-blue EDSM framework~\cite{edsm}. Models are generated by converting the problem into SAT and using state-of-the-art SAT solvers to get an optimal solution. 
%

To avoid over-generalisation in the absence of labelled data, the EDSM algorithm was~improved to incorporate inherent temporal behaviour in the models~\cite{state_merge,Walkinshaw:2007:RES:1339262.1339495}. 
Models are checked against~LTL properties to validate state merges as they are encountered. In an attempt to model and verify software systems~\cite{reflexion_wasim}, state machines describing software behaviour are generated by checking a hypothetical, manually drawn model against the code. The user specifies a set of states and state invariants which are translated into relevant pre and post conditions in the code. State merge based algorithms do not focus on producing the most succinct models but rather produce a good enough approximation that conforms to the trace~\cite{exact_fsm}.

SAT-based approaches to model generation have thus gained popularity due to
their ability to produce exact state
machines~\cite{efsm_state_merge,exact_fsm}.  Similarly, several algorithms
have been developed that use SAT together with state merge to generate
automata from positive and negative
traces~\cite{model_SAT,Buzhinsky2017ModularPM,Heule2013}.  In general, these
methods work by first representing the problem using Boolean variables and
generating a Boolean formula that constrains them.  A~SAT checker then
generates a hypothesis model, which is verified using LTL properties of the
system.
%
%
The SAT-based approach has been put to practice to construct plant models as
Moore machines using behavioural instances, with LTL properties used as
constraints~\cite{model_SAT,Buzhinsky2017ModularPM}.

A classic automata learning technique, Angluin's L* algorithm~\cite{lstar},
employs a series of equivalence and membership queries to an oracle, the
results of which are used to construct the automaton.  When the trace does
not explicitly contain transition predicates, L* fails to learn behaviour
seen in the data.  The absence of an oracle often restricts the use of this
algorithm for abstracting large systems.

\section{Conclusion and Prospects}

In this paper we have outlined a novel scalable program synthesis based
approach to learn models from long execution traces.  The models produced
are concise and accurately represent the system's behaviour.  Our approach
can handle traces that contain more than just Boolean events by synthesising
expressions for system variable state predicates.  We have compared our
approach with state-merge algorithms for a range of benchmarks and evaluated
the scalability of our algorithm.  Our abstract models have several potential
applications: they can summarise which aspects of system behaviour have been
covered by a suite of tests, they provide starting points for model-based
test generation, perhaps to close coverage holes, and they could be used as
candidate inductive invariants~\cite{mc-book}, which, in turn, could be used
to prove properties of the system.

Going forward, we wish to look at these applications and address
the question of how to efficiently exercise the system to produce relevant
traces.  We~are particularly interested in its utility in invariant
synthesis for property verification using the models as candidate invariants
in the inductive invariant refinement loop.

Although we demonstrate our approach on systems given as virtual
platforms, we wish to explore its value in other domains as well.

\section{Acknowledgements}

This research was funded by the Semiconductor Research
Corporation,~Task 2707.001, and Balliol College, Jason Hu~scholarship.  We
thank Daniel Bristot for his help with the RT~Linux Kernel.

\bibliographystyle{IEEEtran}
\bibliography{arXiv}

\end{document}